\documentclass[10pt]{iopart}
\usepackage{iopams}
\usepackage{graphicx}

\newcommand{\be}{\begin{equation}}
\newcommand{\ee}{\end{equation}}
\newcommand{\bea}{\begin{eqnarray}}
\newcommand{\eea}{\end{eqnarray}}
\newcommand{\bse}{\begin{subequations}}
\newcommand{\ese}{\end{subequations}}

\newcommand{\F}{{\mathcal F}}
\newcommand{\Fo}{{\mathcal F}_o}

\newcommand{\ve}{\varepsilon}
\newcommand{\md}{\mathrm{d}}

\newcommand{\phim}{\phi_\mathrm{m}}
\newcommand{\phis}{\phi_\mathrm{s}}

\begin{document}

\title{All-sky search of NAUTILUS data}

\newcounter{affilcount}
\newcommand{\makeaffil}[1]{\addtocounter{affilcount}{1}\edef#1{$^{\arabic{affilcount}}$}}
\newcommand{\affand}{$^,$}

\makeaffil{\AH}
\makeaffil{\CA}
\makeaffil{\DP}
\makeaffil{\IF}
\makeaffil{\IMP}
\makeaffil{\ITP}
\makeaffil{\INF}
\makeaffil{\INR}
\makeaffil{\URL}
\makeaffil{\URF}
\makeaffil{\URR}
\makeaffil{\ING}
\makeaffil{\UA}
\makeaffil{\LNG}
\makeaffil{\UG}

\author{
P~Astone\INR,
M~Bassan\URR,
P~Bonifazi\IF,
K~M~Borkowski\CA,
R~J~Budzy\'nski\DP,
A~Chincarini\ING,
E~Coccia\URR$^,$\LNG,
S~D'Antonio\URR,
M~Di Paolo Emilio\UA$^,$\LNG,
V~Fafone\INF,
S~Frasca\URL,
S~Foffa\UG,
G~Giordano\INF,
P~Jaranowski\ITP,
W~Kondracki\IMP,
A~Kr\'olak\IMP\footnote[8]{krolak@impan.gov.pl},
M~Maggiore\UG,
A~Marini\INF,
Y~Minenkov\URR,
I~Modena\URR,
G~Modestino\INF,
A~Moleti\URR,
G~V~Pallottino\URL,
C~Palomba\INR,
R~Parodi\ING,
M~Pietka\ITP,
G~Pizzella\URF,
H~J~Pletsch\AH,
L~Quintieri\INF,
F~Ricci\URL,
A~Rocchi\URR,
F~Ronga\INF,
R~Sturani\UG,
R~Terenzi\IF,
R~Vaccarone\ING, and
M~Visco\IF}

\address{\AH Albert-Einstein-Institut, Max-Planck-Institut f\"ur Gravitationsphysik, Hannover, Germany}
\address{\CA Centre of Astronomy, Nicolaus Copernicus University, Toru\'n, Poland}
\address{\DP Department of Physics, Warsaw University, Warsaw, Poland}
\address{\UG Dep. de Phys. $\mathrm{Th\acute{e}orique,~Universit\acute{e}~de~Gen\grave{e}ve,~Gen\grave{e}ve,~Switzerland}$}
\address{\IF IFSI-CNR and INFN, Rome, Italy}
\address{\LNG INFN, Laboratori Nazionali del Gran Sasso, Assergi, L'Aquila, Italy}
\address{\IMP Institute of Mathematics, Polish Academy of Sciences, Warsaw, Poland }
\address{\ITP Faculty of Physics, University of Bia{\l}ystok, Bia{\l}ystok, Poland}
\address{\INF Istituto Nazionale di Fisica Nucleare INFN, Frascati, Italy}
\address{\ING Istituto Nazionale di Fisica Nucleare INFN, Genova, Italy}
\address{\INR Istituto Nazionale di Fisica Nucleare INFN, Rome, Italy}
\address{\UA  $\mathrm{Universit\grave{a}}$ dell'Aquila, Italy}
\address{\URL University of Rome ``La Sapienza" and INFN, Rome, Italy}
\address{\URF University of Rome ``Tor Vergata" and INFN, Frascati, Italy}
\address{\URR University of Rome ``Tor Vergata" and INFN, Rome II, Italy}


\begin{abstract}
A search for periodic gravitational-wave signals from isolated neutron stars in the
NAUTILUS detector data is presented. We have analyzed half a year of data over
the frequency band $\langle922.2;\,923.2\rangle$\,Hz, the spindown
range $\langle-1.463\times10^{-8};\,0\rangle$\,Hz/s and over the entire sky.
We have divided the data into 2 day stretches and we have analyzed each stretch coherently
using matched filtering. We have imposed a low threshold for the optimal detection statistic
to obtain a set of candidates that are further examined for coincidences among various data
stretches. For some candidates we have also investigated the change of the signal-to-noise
ratio when we increase the observation time from two to four days. Our analysis has not revealed
any gravitational-wave signals. Therefore we have imposed upper limits on the dimensionless
gravitational-wave amplitude over the parameter space that we have searched.
Depending on frequency, our upper limit ranges from $3.4 \times 10^{-23}$ to $1.3 \times 10^{-22}$.
We have attempted a statistical verification of the hypotheses leading to our conclusions.
We estimate that our upper limit is accurate to within $18$\%.

\end{abstract}


\submitto{\CQG}

\maketitle

\section{Introduction}
We present results of the search of the NAUTILUS (a resonant bar detector \cite{Naut})
data for periodic gravitational-wave (GW) signals.
We search the band $\langle922.2;\,923.2\rangle$\,Hz and the spindown range $\langle-1.463\times10^{-8};\,0\rangle$\,Hz/s over the entire sky.
We analyze NAUTILUS data collected in the year 2001. We divide the data into stretches of 2
sidereal days. Each stretch of data is analyzed coherently using matched filtering in
the form of the $\F$-statistic \cite{JKS98,JKLivRev}. We have analyzed slightly more than half a year
of data.

Previous analysis of bar detector data for periodic GW signals were
the search of the galactic center and the globular cluster 47 Tucanae
with the ALEGRO detector \cite{Mau00}, the search of the galactic center using the EXPLORER detector data \cite{Astone02}, and an all-sky search
using the EXPLORER data \cite{ul1,ul2}. LIGO (Laser Gravitational Wave Observatory) data
was searched for known pulsars \cite{LIGO04,LIGOl05,LIGOl07}, over the entire sky \cite{LIGO07}
using the coherent method and over all the sky using incoherent methods \cite{LIGO05,LIGO08}.
Currently LIGO data are analyzed over the entire sky by the Einstein@Home project \cite{eah08}.

In section \ref{Sec:Met} we present the data analysis methods used in our search.
In section \ref{Sec:Pro} we outline our search procedure. In section \ref{Sec:Can} we discuss
the analysis of the candidates.  This analysis consists of two parts: the first part is the search for coincidences between the candidates each obtained from a different stretch of data and the second part
is an investigation of the increase of the signal-to-noise ratio of candidates when we increase
the observation time from two to four days. In section \ref{Sec:UL} we impose upper limits on
amplitudes of the gravitational waves in the parameter space that we have searched.

\section{Data analysis methods}
\label{Sec:Met}
In order to search for gravitational waves from long lived periodic sources we have
used the maximum likelihood (ML) method. For the case of Gaussian noise the ML method
consists of linearly filtering the data with a template matched to the signal that we
are searching for. The main complication of the matched filtering is that the signal
depends on several unknown parameters. This requires evaluation of the likelihood function
over a large parameter space. In order to minimize the computation time we use several
data analysis tools. {\em Firstly} we find the maximum likelihood estimators of some parameters
(4 in the case of a GW signal from a rotating neutron star that we are searching for) in a
closed analytic form, thereby reducing the dimensionality of the parameter space that we have to search.
The likelihood function over the reduced parameter space is called  the $\F$-statistic and it
is derived in \cite{JKS98}. {\em Secondly} we analyze data of length equal to an integer multiple of
a sidereal day. This leads to a considerable simplification of the $\F$-statistic and consequently
reduces the number of numerical operations to evaluate it.
The $\F$-statistic for observation time equal to an integer number of sidereal days is given
in section III of \cite{puls4}.
{\em Thirdly} we use optimal numerical algorithms, in particular the Fast Fourier Transform
(FFT) in order to calculate the $\F$-statistic efficiently.
{\em Fourthly} we minimize the number of $\F$-statistic calculations
over the parameter space by solving a covering problem for this space \cite{Con,Prix07}.
Let us explain the latter two data analysis tools in more detail. The response of a bar
detector to a gravitational-wave signal from a spinning neutron star is summarized
in section 2.1 of \cite{ul2}.

{\em Fast Fourier Transform.}  Estimates have shown \cite{BCCS98,JK00} that for the bandwidth and the spindown range that we search we need to take into account in our
templates only one  spindown parameter in order to match the signal. Consequently
the phase modulation function $\phi(t)$ of the waveform is given by
\be
\label{pha3}
\phi(t) = \omega_0 t + \omega_1 t^2
+ ( \omega_0 + 2 \omega_1 t )\frac{{\bf n}_0\cdot{\bf r}_{\rm d}(t)}{c},
\ee
where $\omega_0$ is angular frequency and $\omega_1$ is the spindown
parameter, $\mathbf{n}_0$ is the constant unit vector
in the direction of the star in the Solar System Barycenter (SSB) reference frame
(it depends on the right ascension $\alpha$ and the declination $\delta$ of the source),
and ${\bf r}_{\rm d}$ is the vector joining the SSB with the detector and $c$ is the speed
of light. The detection statistic $\F$ involves two integrals of the form
\be
\label{eq:Fab}
F = \int^{T_0}_0 x(t) \, a(t) \, e^{-i\phi(t)}
\,\md t,
\ee
where $x(t)$ is the data stream, $a(t)$ is the amplitude modulation function that depends on $\delta$
and $\alpha$.
The above integral is not a Fourier transform because the frequency
$\omega_0$ in the phase multiplies the term ${\bf n}_0\cdot{\bf r}_{\rm d}(t)$
which is a non linear function of time. In order to convert the integral
into a Fourier transform we introduce the following interpolation procedure.
The phase $\phi(t)$ [Equation\ (\ref{pha3})] can be written as
\be
\phi(t) = \omega_0[t + \phim(t)] + \phis(t),
\ee
where
\bea
\label{pham}
\phim(t) &:= \frac{{\bf n}_0\cdot{\bf r}_{\rm d}(t)}{c},
\\
\label{phas}
\phis(t) &:= \omega_1 t^2
+ 2 \frac{{\bf n}_0\cdot{\bf r}_{\rm d}(t)}{c} \omega_1 t.
\eea
The functions $\phim(t)$ and $\phis(t)$
do not depend on the angular frequency $\omega_0$.
We can write the integral (\ref{eq:Fab}) as
\be
\label{eq:Far}
F = \int^{T_0}_0 x(t) \, a(t) \, e^{-i\phis(t)}
\exp\big\{-i\omega_0[t + \phim(t)]\big\}\,\md t.
\ee
We see that the integral (\ref{eq:Far}) can be
interpreted as a Fourier transform (and computed efficiently with
an FFT), if $\phi_{\mathrm m}=0$. In order to convert equation\ (\ref{eq:Far})
to a Fourier transform we introduce a new time variable $t_b$,
so called {\em barycentric time} \cite{S91,JKS98},
\be
\label{eq:Bt}
t_b := t + \phim(t).
\ee
In the new time coordinate the integral (\ref{eq:Far})
is approximately given by (see Ref.\ \cite{JKS98}, Sec.\ IIID)
\be
\label{ia}
F \cong \int^{T_0}_{0} x[t(t_b)] a[t(t_b)]
e^{-i\phis[t( t_b)]} e^{-i\omega_0 t_b}\,\md t_b.
\ee
Thus in order to compute the integral (\ref{eq:Far}),
we first multiply the data $x(t)$ by the function $a(t)\,\exp[-i\phis(t)]$
for each set of the parameters $\omega_1, \delta, \alpha$
and then resample the resulting function according to equation\ (\ref{eq:Bt}).
At the end we perform the FFT.

{\em The covering problem.} The covering problem is to find the minimum number
of templates in the parameter space \cite{Prix07}, so that the fractional loss in signal to ratio is not less than $1 - MM$ ($MM$ is the {minimal match} parameter introduced by Owen \cite{owen-96}). In order to solve the covering problem we introduce a useful approximate model
of the gravitational-wave signal from a rotating neutron star.
The model relies on (i) neglecting all spindowns in the phase modulation
due to motion of the detector with respect to the SSB;
and (ii) discarding the component of the vector ${\bf r}_{\rm d}$
(connecting the SSB and the detector) which is perpendicular to the ecliptic plane.
These approximations lead to the following phase model of the signal:
\be
\label{philin}
\phi_\mathrm{lin}(t) = \omega_0 t + \omega_1 t^2
+ \alpha_1 \mu_1(t) + \alpha_2 \mu_2(t),
\ee
where $\alpha_1$ and $\alpha_2$ are new constant parameters,
\bea
\alpha_1 &:= \omega_0 (\sin\alpha\cos\delta\cos\ve + \sin\delta\sin\ve),
\\[1ex]
\alpha_2 &:= \omega_0 \cos\alpha\cos\delta,
\eea
where $\ve$ is the obliquity of the ecliptic
and where $\mu_1(t)$ and $\mu_2(t)$ are known functions of time,
\bea
\mu_1(t) &:= R^y_\mathrm{ES}(t) + R^y_\mathrm{E}(t)\cos\ve,
\\[1ex]
\mu_2(t) &:= R^x_\mathrm{ES}(t) + R^x_\mathrm{E}(t).
\eea
$R^x_\mathrm{ES}$ is the $x$-component of the vector joining the
center of Earth and the SSB, and $R^x_\mathrm{E}$ is the $x$-component of the vector joining
the center of Earth and the detector. $R^y_\mathrm{ES}(t)$ and $R^y_\mathrm{E}(t)$
are the corresponding $y$-components.
We also neglect the slowly varying modulation of the signal's amplitude,
so finally we approximate the whole signal $h(t)$ by
\be
\label{eq:reslin}
h(t) = A_0\cos\big(\phi_\mathrm{lin}(t)+ \phi_0\big),
\ee
where $A_0$ and $\phi_0$ are the constant amplitude and initial phase, respectively.
The above signal model is called {\em linear}
because it has the property that its phase given by equation\,(\ref{philin})
is a linear function of the parameters. We have shown \cite{JK99} that the above model is a good approximation to the accurate response of the detector to the GW signal
in the sense that the Fisher matrix for the linear model reproduces well the Fisher matrix
for the accurate model. Thus whenever a Fisher matrix is needed we can use the Fisher matrix
for the linear model as an approximation to the Fisher matrix for the accurate model.
The great advantage of the linear model is that components of its Fisher matrix are constant,
independent of the values of the parameters. In order to solve the covering problem for
the parameter space we use the Fisher matrix as a metric on the parameter space. Because
the components of the Fisher matrix are constant the grid is uniform what greatly simplifies
its construction. In our search, as a grid we use the hypercubic lattice \cite{Con}.
However we have an additional constraint. In order to apply the FFT algorithm the nodes of the grid
must coincide with the Fourier frequencies. We have constructed a suitable grid by performing
rotations and dilatations of the original hypercubic lattice. We construct the grid
in the parameters $\omega_0, \omega_1, \alpha_1, \alpha_2$ and then transform it to
parameters $\omega_0, \omega_1, \delta, \alpha$ for which the $\F$-statistic is calculated.

The linear parametrization has one more application. We use it in order to calculate
the threshold for the $\F$-statistic corresponding to a certain false alarm probability.
Namely, using the linear parametrization we divide the parameter space into cells
as explained in \cite{JKS98,JK00}. All the cells are exactly the same and their number
$N_\mathrm{c}$ is easily calculated using the Fisher matrix
(see section IIIB of \cite{JK00}).
The false alarm  probability $\alpha$ is the probability that $\F$ exceeds
threshold $\Fo$ in \emph{one or more} cells and is given by
\be
\label{FP}
\alpha = 1 - \big[1 - P_F(\Fo)\big]^{N_\mathrm{c}},
\ee
where $P_F$ is the false alarm probability for a single cell.

\section{Search procedure}
\label{Sec:Pro}
\begin{figure}[t]
\centering
\includegraphics[width=12cm]{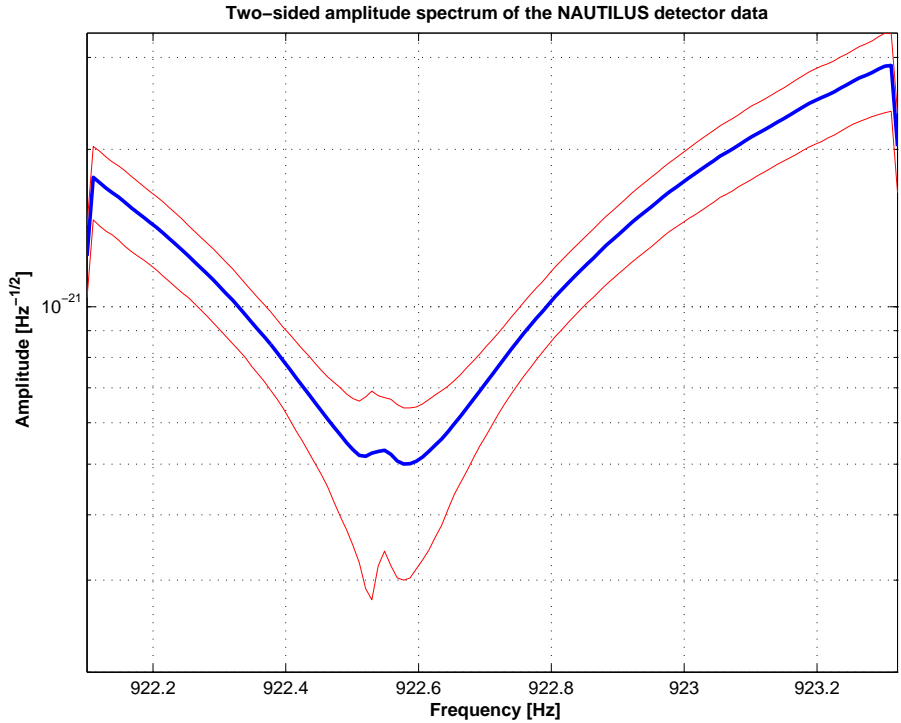}
\caption{\label{fig:spec} Estimation of the two-sided amplitude spectrum of the NAUTILUS data
in the year 2001 and the rms error of the estimate. The thick line shows the estimate
and the two thin lines correspond to the 1 $\sigma$ error.}
\end{figure}

We have searched the data collected by the NAUTILUS detector in the year 2001.
The bandwidth of $\langle922.2;\,923.2\rangle$\,Hz, where the detector
is most sensitive, has been analyzed. We have divided the data into stretches
which span two sidereal days.
We have assumed a minimum pulsar spindown age $\tau_{min}$ equal to $1000$\,yrs
and so we have searched the negative frequency time derivatives
in the range of $\langle-1.463\times10^{-8};\,0\rangle$\,Hz/s.  For this $\tau_{min}$
and two days of the observation time it is sufficient to include only one spindown
in the phase \cite{BCCS98,JK00}. Each two day sequence was analyzed coherently using the $\F$-statistic.
We have used the constrained hypercubic grid as explained in the previous section.
For the grid construction we have assumed the minimal match parameter
$MM = \sqrt{3}/2$ \cite{owen-96}. Using this
minimal-match value our grid consists of around $3.1\times10^{13}$ grid points
($2^{19}$ frequency bins, $\sim 10^3$ spindowns, $\sim 6\times10^4$ sky positions).
The threshold on 2$\F$ corresponding to $1\%$ false alarm probability has been calculated
using Equation~(\ref{FP}) and is around 72. In order to compensate
the loss of signal-to-noise ratio (SNR) due to the discreteness of the grid, imperfect
templates and numerical approximation in evaluation of the $\F$-statistic (resampling procedure)
we have adopted two lower thresholds on 2$\F$ equal to $40$
and $50$. We have registered parameters of all templates which crossed the threshold
of $40$. For threshold crossings of $50$ we have performed a
verification procedure. The verification procedure consisted of calculating
the $\F$-statistic for the template parameters of the candidate
using a four day stretch of data involving the original two day stretch. For a
true gravitational-wave signal by this procedure one would expect an increase of signal-to-noise ratio
by $\sqrt{2}$. In total, we have analyzed 93 two day data stretches.
In Figure~\ref{fig:spec} we have presented the two-sided amplitude spectrum of the NAUTILUS
detector data that we have analyzed. The spectrum was obtained in the following way.
We have estimated the power spectrum density in each of the 93 two day data sequences
and then we have taken the square root of the average of the 93 power spectra.
We see that the best sensitivity is around $5\times 10^{-22}\,$Hz$^{-1/2}$.
Moreover we have obtained the rms error of our power spectrum estimate by calculating the
variance from the estimates of the spectra of in each of the 93 data segments. The relative 1 $\sigma$
error in the amplitude power spectrum is around 18\%.

During the search we have obtained $537\,665\,380$ candidates above the $2\F$-threshold
of $40$ and $9\,038\,817$ above the $2\F$-threshold of $50$.

\section{Analysis of the candidates}
\label{Sec:Can}
\subsection{Signal-to-noise ratio of the candidates}
\begin{figure}[t]
\centering
\includegraphics[width=12cm]{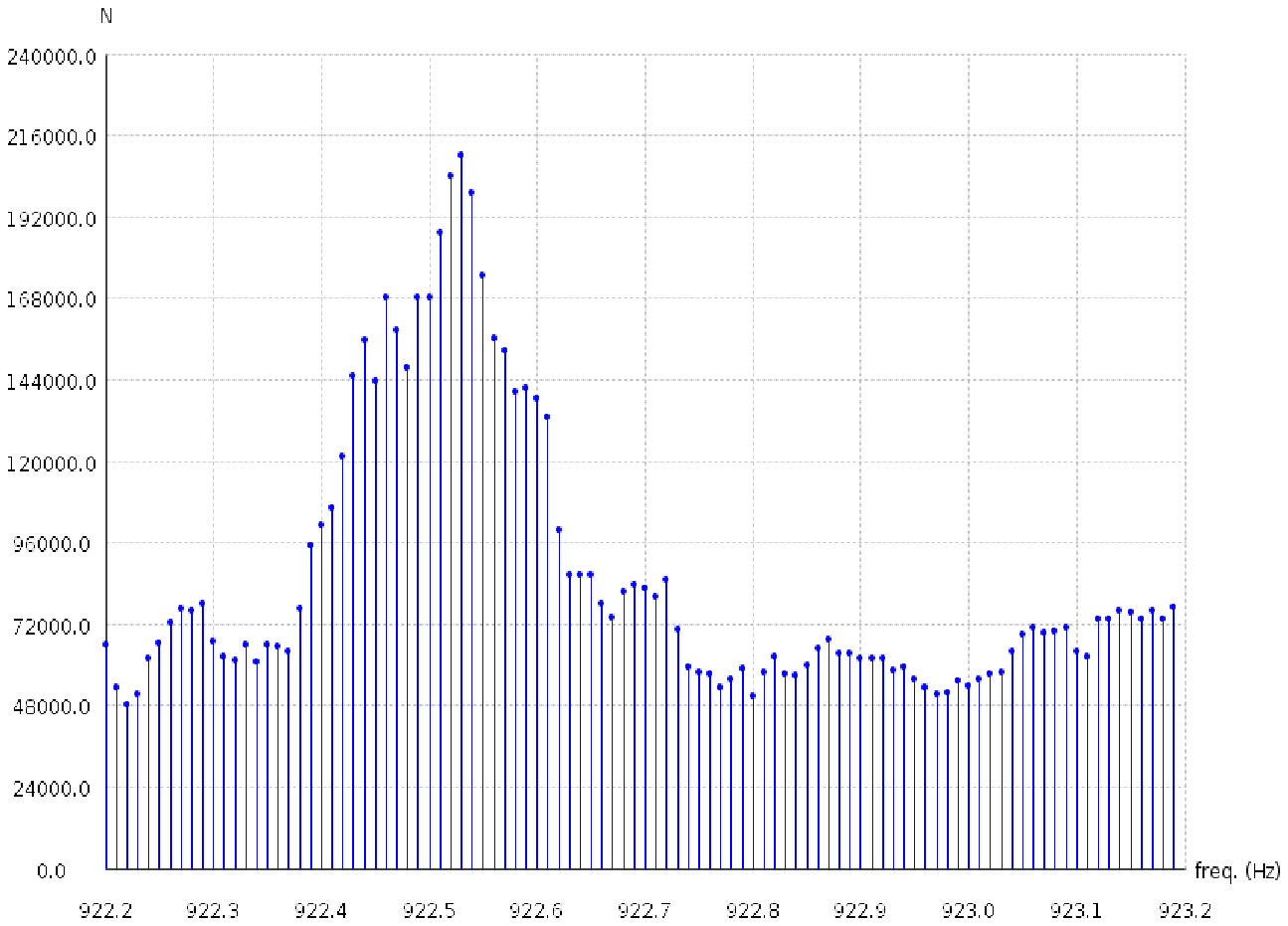}
\caption{\label{fig:freqhist} Histogram of the frequencies of candidates
obtained in the search of all 93 two day data stretches above a $2\F$-threshold of 50.}
\end{figure}

In Figure~\ref{fig:freqhist} we have plotted a histogram of the frequencies
of all the candidates above the $2\F$-threshold of 50. The histogram shows an excess
of candidates in the frequency band of $\langle922.4;\,922.6\rangle$\,Hz. This excess is a
result of the presence of a periodic interference in the data that appears as
a series of harmonics in the bandwidth of the detector. One of the harmonics is located
in the subband $\langle922.2;\,923.2\rangle$\,Hz. The effect of the harmonic
is visible in our estimate of the spectrum (Figure~\ref{fig:spec}) and appears
as a bump in the band $\langle922.5;\,922.6\rangle$\,Hz.

\begin{figure}[t]
\centering
\includegraphics[width=12cm]{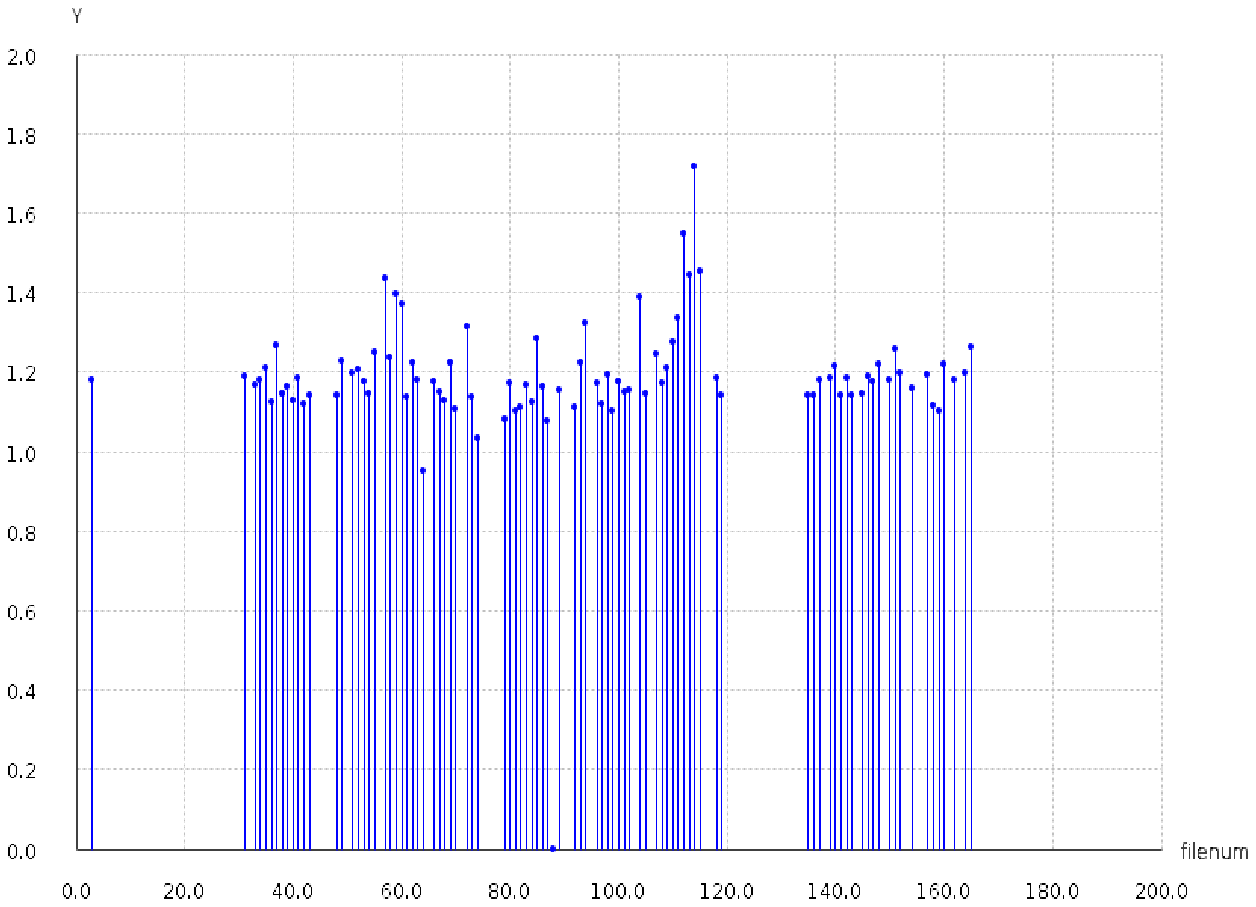}
\caption{\label{fig:SNRincr} Highest increase (vertical axis) in signal-to-noise ratio
for candidates in each of the 93 data stretches analyzed. The two day stretches of
Nautilus 2001 data are numbered form 1 to 182. The missing lines in the plot indicate
that the corresponding data stretch was not analyzed.}
\end{figure}

As a first step in the candidate analysis we have calculated the increase in signal-to-noise ratio
when we increase the observation time from two to four days. This has been done for
all the candidates above the threshold $2\F = 50$.
Figure~\ref{fig:SNRincr} shows the highest increase in SNR for candidates when going
from a two day data stretch to the four day one.
The maximum is calculated for each of the 93 data stretches analyzed.
We see that typically the highest gain in the signal-to-noise is $1.2$. This should be compared
with the theoretical gain of $\sqrt{2}$ of SNR when we increase the observation time
by a factor of 2. The periodic interference present in the data to which we attribute
these maximum SNR increases does not gives a higher increase
of the SNR because its frequency changes erratically over the observation time
of days and it cannot reproduce the Doppler shift of a real GW signal
modulated by detector motion with respect to the SSB. Assuming that the two day sequence is independent of the four day sequence
we could perform the $F$-test that consists of calculation of the ratio $F$ of the $\F$-statistic
for 4 days observation time and the $\F$-statistic for 2 days observation time. Taking as the null
hypothesis for the test that data is only Gaussian noise the $2\F$-statistic
has the central $\chi^2$ distribution with 4-degrees of freedom and the ratio $F$ has Fisher-Snedeckor distribution $F(4,4)$. The typical highest value of $F$ for a given data segment is around $1.5$
The probability of $F$ crossing the threshold $1.5$ is around $37$\%. This would give
a high confidence that data is noise only. Unfortunately this is only a crude approximation
because the two day sequence is contained in the four day one and the assumption of independence
of the two $\F$-statistic is not fulfilled.

\subsection{Coincidences among the candidates from different data stretches}
Candidates from different data stretches are considered {\it coincident} if they
cluster closely together in the four-dimensional parameter
space~$(\omega_0, \omega_1, \delta, \alpha)$. We employ the
clustering method described in~\cite{eah08}, which uses a grid
of ``coincidence cells". This method will reliably detect strong signals
which would produce candidates with closely-matched parameters in
many of the different data stretches.

In a first step, the frequency value of each candidate above the threshold
of $2\F=40$ is shifted to the same fiducial time: the GPS start
time of the earliest ($j=1$) stretch, $t_{\rm fiducial} = t_1=662\,547\,735.9988098\,{\rm s}$.
Defining $T_0$ to be the time span of two sidereal days, the frequencies of the
candidates are shifted to~$t_{\rm fiducial}$ via
\begin{equation}
  \omega_0(t_{\rm fiducial}) = \omega_0(t_j) + ( j - 1 )\,2 \omega_1T_0 ,
\end{equation}
where $t_j$ is the starting time of the $j$'th data stretch,
given by $ t_j =  t_{\rm fiducial} + ( j - 1 )\,T_0$.

To find coincidences, a grid of cells is constructed such that the cells
are rectangular in the coordinates $(\omega_0, \omega_1, \delta, \alpha \cos \delta)$.
The dimensions of the cells are adapted to the parameter space search.
Thus, the cells are constructed to be as small as possible to reduce the probability
of coincidences due to false alarms.  However, since each of the 93~different data stretches
uses a slightly different parameter space grid, the coincidence cells must be chosen
to be large enough that the candidates from a source (which
would appear at slightly different points in parameter space in each
of the 93~data stretches) would still lie in the same coincidence cell.
As a conservative choice we use cell sizes in $\omega_0$ of $5.8\times10^{-4}\,{\rm Hz}$,
in $\omega_1$ of $2.08\times10^{-11}\,{\rm Hz\,s^{-1}}$, and
an isotropic cell grid in the sky with equatorial spacing of $0.028\,{\rm rad}$.
Each candidate event is assigned to a
particular cell.  In cases where two or more candidate events from the
same data stretch~$j$ fall into the same cell, only the candidate
having the largest value of $2\F$ is retained in the cell.  Then
the number of candidate events per cell coming from distinct data
stretches is counted.

From the 93~different data stretches, this coincidence method
found that we get candidates which appear consistently
in no more than 4 data stretches uniformly over the search bandwidth, where
there are no instrumental interferences.
This is the background of the number of coincidences.
We would like to test the null hypothesis that the coincidences are result of the noise only.
Let us assume that the parameter space is divided into $N_{cell}$ independent coincidence cells,
the candidate events are independent and the probability for a candidate
event to fall into any given coincidence cell is $1 = 1/N_{cell}$.
Thus probability $\epsilon$ that a given coincidence cell is populated with one or more candidate event
is given by
\be
\epsilon = 1 - (1 - \frac{1}{N_{cell}})^{\varepsilon_{seg}},
\ee
where $\varepsilon_{seg}$ is the number of candidate events per data segment.
The probability $p_F$ that any given coincidence cell contains candidate events from
${\mathcal C}_{max}$ or more distinct data segments is given by a binomial distribution
\be
p_F = \sum^{N_{seg}}_{n = {\mathcal C}_{max}} {N_{seg} \choose n}\epsilon^n(1 - \epsilon)^{N_{seg}-n}.
\ee
Finally the probability $P_F$ that there is ${\mathcal C}_{max}$ or more coincidences in one or
more of the $N_{cell}$ cells is
\be
\label{eq:FAcoin}
P_F = 1 - (1 - p_F)^{N_{cell}}.
\ee
The expected number of cells with ${\mathcal C}_{max}$ or more coincidences is given by
\be
N_F = N_{cell}\,p_F.
\ee
In our case the number of cells is given by $N_{cell} = 5.9 \times 10^{10}$,
the number of data segments is $N_{seg} = 93$, and the number of candidates
per data segment is $\varepsilon_{seg} = 5.8 \times 10^6$. From Equation~(\ref{eq:FAcoin})
we find that the probability of finding ${\mathcal C}_{max} = 4$ or more coincident
candidates is almost one. Thus for the background coincidences we can accept the null
hypothesis that they are from noise only with a high confidence. Over the bandwidth
$\langle922.4;\,922.6\rangle$\,Hz we find an excess of coincidences with the maximum of 8
coincidences. By Equation~(\ref{eq:FAcoin}), the false alarm probability associated with
with 8 or more coincidences is of the order of $10^{-11}$ and thus they cannot be attributed
to noise. We consider these coincidences to be due to the periodic interference present in the
data.

\section{Upper limits}
\label{Sec:UL}
\begin{figure}[t]
\centering
\includegraphics[width=12cm]{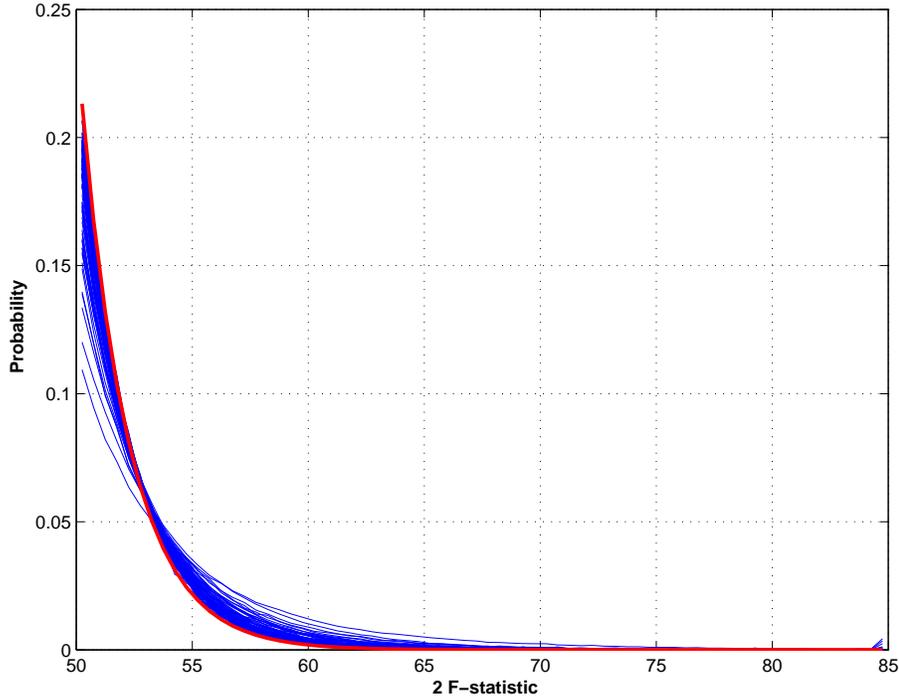}
\caption{\label{fig:TrigHist} Probability distribution of $2\F$-statistic values of the
candidates. The light lines are obtained from histograms of the $2\F$ values of each data segment.
The thick line represents the theoretical central $\chi^2$ distribution with 4 degrees of freedom.}
\end{figure}

Our verification procedure consisting of coincidences among the candidates from
distinct data segments and an analysis
of the increase of signal-to-noise ratio presented in section \ref{Sec:Can} did
not produce convincing evidence of a gravitational-wave signal.
We therefore proceeded to estimate the upper limits for the amplitudes of
the gravitational-wave signals in the parameter space that we have searched.
Detection of a signal is signified by a large value of the
$\F$-statistic that is unlikely to arise from the noise-only
distribution. If instead the value of $\F$ is consistent with pure
noise with high probability we can place an upper limit on the
strength of the signal. One way of doing this is to take the
loudest event obtained in the search and solve the equation
\begin{equation}
\label{UL}
{\cal P} = P_D(\rho_\mathrm{ul},\F_\mathrm{loudest})
\end{equation}
for signal-to-noise ratio $\rho_\mathrm{ul}$,
where $P_D$ is the detection probability,
$\F_\mathrm{loudest}$ is the value of the $\F$-statistic
corresponding to the loudest event,
and ${\cal P}$ is a chosen confidence.
Then $\rho_\mathrm{ul}$ is the desired upper limit with confidence ${\cal P}$.
We can also obtain an upper limit $\rho_\mathrm{ul}$ with confidence ${\cal P}$
for several independent searches from the equation
\be
\label{ULL}
{\cal P} =  1 - \prod^L_{s=1}
\left[1 - P_D(\rho_\mathrm{ul},\F_{\mathrm{loudest}\,s})\right],
\ee
where $\F_{\mathrm{loudest}\,s}$ is the threshold corresponding to the loudest
event in $s$'th search and $L$ is the number of searches.
Here ${\cal P}$ is the probability that a signal with signal-to-noise ratio
$\rho_\mathrm{ul}$ crosses the threshold $\F_{\mathrm{loudest}\,s}$ in at least one of the
$L$ independent searches. To calculate $\rho_\mathrm{ul}$ we assume that the
data have a Gaussian distribution and consequently the probability of detection
$P_D$ has a non-central $\chi^2$ distribution with 4 degrees of freedom
and the noncentrality parameter equal to $\rho^2_\mathrm{ul}$ . We have investigated
this assumption by obtaining histograms of the $2\F$-statistic values of the candidates
and comparing them to the central $\chi^2$ distribution with 4 degrees of freedom.
The result is shown in Figure~\ref{fig:TrigHist}. There is an overall qualitative
agreement of candidates distributions with the theoretical one. However, the candidates distributions
do not pass a goodness-of-fit test for a $\chi^2$ distribution at the significance level of 5\%.

\begin{figure}[t]
\centering
\includegraphics[width=12cm]{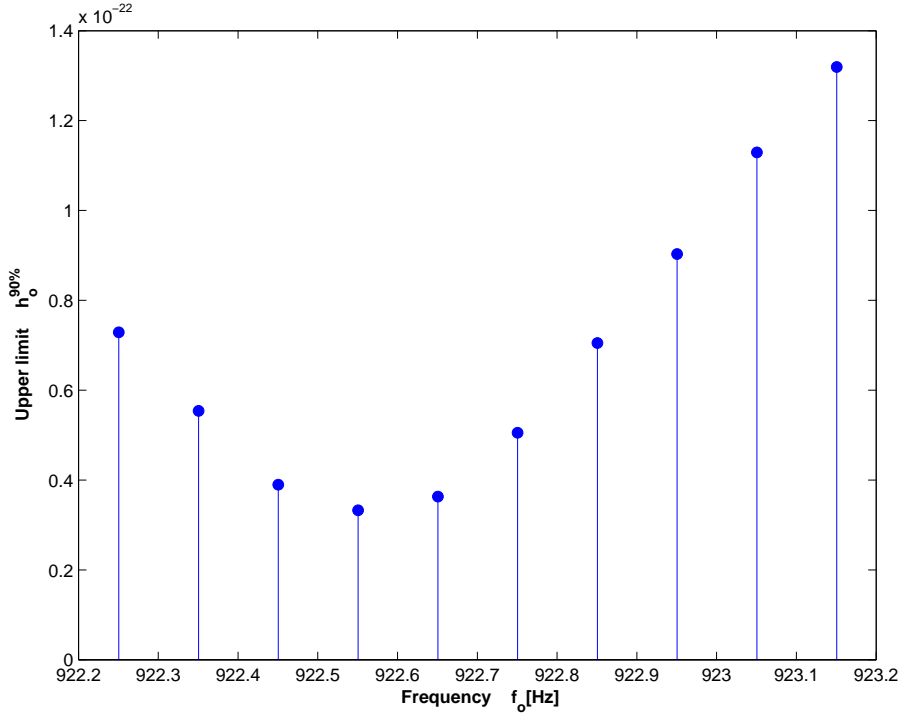}
\caption{\label{fig:UpLim} Upper limits based on the loudest candidate for 0.1 Hz
frequency bands over the 1 Hz bandwidth searched.}
\end{figure}

In order to translate our upper limit on the SNR
into the upper limit on the gravitational-wave amplitude, we use the
Equation~(93) of \cite{JKS98} for signal-to-noise ratio of a GW signal
from a spinning neutron star averaged over the source position and orientation.
Thus $h_\mathrm{ul}$ and $\rho_\mathrm{ul}$ are related by the following formula:
\be
\label{eq:ul}
h_\mathrm{ul}(f) = \frac{5}{2}\sqrt{\frac{S(f)}{T_0}}\rho_\mathrm{ul},
\ee
where $S(f)$ is one-sided spectral density at frequency $f$.
We have used Equations~(\ref{ULL}) and (\ref{eq:ul}) to obtain
upper limits in $0.1$ Hz bands over the bandwidth $\langle922.2;\,923.2\rangle$\,Hz that we have searched.
The upper limit results are presented in Figure~\ref{fig:UpLim}.
Assumed Gaussian noise, we have chosen the confidence ${\cal P} = 90\%$ and
we denote the upper limits by $h_o^{90\%}$.
Our best upper limit is equal to $3.4 \times 10^{-23}$ at a frequency of
922.55 Hz. Using our 1 $\sigma$ rms error of the amplitude power spectrum estimate
we reckon that our upper limit has likewise an error of $18$\%.

\section*{Acknowledgments}
We would like to thank Maria Alessandra Papa for making available to us large computing resources of the Albert-Einstein-Institut enabling to complete the data analysis for this paper in a reasonable time.
We would also like to thank the Interdisciplinary Center for Mathematical and Computational Modelling
of Warsaw University for computing time. A Kr\'olak would like to acknowledge the financial support of Istituto Nazionale di Fisica Nucleare (INFN), the INFN Section in University of Rome ``La Sapienza'' and the Max-Planck-Institut f\"ur Gravitationsphysik (Albert-Einstein-Institut). H Pletsch would like to acknowledge the support of
the IMPRS on Gravitational Wave Astronomy.

\end{document}